\documentclass[a4paper]{jpconf}

\usepackage{graphicx}
\usepackage{iopams}
\usepackage{amssymb,amsmath}
\usepackage{bm}
\usepackage{mcite}

\begin{document}
\title{Transport coefficients in neutron star cores  in BHF approach. Comparison of different nucleon potentials}

\author{P S Shternin$^1$, M Baldo$^2$ and H-J Schulze$^2$ }

\address{$^1$ Ioffe Institute, St. Petersburg, Russia}
\address{$^2$ INFN Sez. di Catania, Catania, Italy}

 \ead{pshternin@gmail.com}

\def\apj{ApJ}
\def\mnras{MNRAS}
\def\nat{Nature}
\def\aap{A\&A}
\def\prc{Phys. Rev. {\rm C}}
\def\prb{Phys. Rev. {\rm B}}
\def\prd{Phys. Rev. {\rm D}}

\begin{abstract}
Thermal conductivity and shear viscosity of npe$\mu$ matter in
non-superfluid neutron star cores are considered in the framework
of Brueckner-Hartree-Fock many-body theory. We extend our previous
work (Shternin {\em et al} 2013 {\em PRC} {\bf 88} 065803) by analysing different
nucleon-nucleon potentials and different three-body forces. We
find that the use of different potentials leads up to one order of
magnitude variations in the values of the nucleon contribution to
transport coefficients. The nucleon contribution dominates the
thermal conductivity, but for all considered models the shear
viscosity is dominated by leptons.

\end{abstract}

\section{Introduction}
Transport coefficients serve as an important part of the
microphysical input required for modelling of various
non-equilibrium processes in neutron stars (NSs). For instance,
thermal conductivity is needed to study the NS cooling (especially
at early stages) and thermal wave propagation from internal
heating sources \cite{Yakovlev2001physrep}. Viscosity coefficients
regulate the damping and stability of the oscillation modes within
the stars \cite{Haskell2015IJMPE}. In these proceedings we
consider the nucleon contribution to the thermal conductivity
$\kappa$ and the shear viscosity $\eta$  in non-superfluid NS
cores in equilibrium with respect to beta-processes with the
simplest npe$\mu$ composition. Transport coefficients of a mixture
are governed by the collisions between the species.
% of the strongly-interacting matter in
%equilibrium state with respect to $\beta$ processes.
%Transport coefficients of the npe$\mu$ matter are usually obtained
%in the framework of the multicomponent Fermi-liquid theory.
In NS cores conditions, the lepton and nucleon sub-systems
decouple and these sectors can be considered separately (e.g.,
\cite{FlowersItoh1979ApJ}).

%Here we focus on the nucleon contribution to the transport
%coefficients.
In \cite{Shternin2013PhRvC}, $\kappa$ and $\eta$ were calculated
in the non-relativistic Brueckner-Hartree-Fock (BHF) approach for
one particular nucleon-nucleon potential, Argonne v18 (Av18 in
short) supplemented by the phenomenological Urbana IX (UIX in
short) model for three-body forces. It is known, that purely
two-body interactions in non-relativistic calculations can not
correctly reproduce the empirical nuclear matter saturation point
and the inclusion of three-body forces is necessary. Several
models for the three-body forces are available on the market, and
we investigate the model-dependence of the results obtained in
\cite{Shternin2013PhRvC}. In this respect, we employ the same
microscopic potentials as in \cite{Baldo2014PhRvC}, where
appropriate references and more detailed discussion can be found.
Namely, in addition to the Av18+UIX model, we also use the CD-Bonn
two-body potential with the corresponding adjustment of UIX
three-body interaction. Both the Av18+UIX and CD-Bonn+UIX
combinations reproduce the correct saturation point. In addition,
we employ the three-body force obtained within the meson-nucleon
theory, where the same meson-exchange parameters are used for two-
and three- body interactions.
 As in \cite{Baldo2014PhRvC}, we use this model constructed for Argonne
v18 potential and denote it as Av18+TBF.

\section{Formalism}\label{sec:form}
The contribution of each strongly degenerate species' in the
mixture can be written as (we omit the species index, for brevity)
\begin{equation}\label{eq:kappa_eta}
  \kappa = C_\kappa \frac{\pi^2 k_{\rm B}^2 T n}{3 m^*} \tau,\quad
  \eta = C_\eta \frac{ p_{{\rm F}}^2 n}{5 m^*} \tau,
\end{equation}
where $T$ is the temperature, $p_{\rm F}$ is the Fermi momentum,
$n$ is the number density, $m^*$ is the effective mass on the
Fermi surface, $k_{\rm B}$ is the Boltzmann constant, $\tau$ is
the characteristic relaxation time and $C_{\kappa,\eta}$ are
numerical constants.
%A separation between $C_{\kappa,\eta}$ and
%$\tau$ is somewhat arbitrary and will be made explicit below.
The latter values are found as a solution of the system of
transport equations for the multicomponent Fermi-liquid.
% and depend
%on the rates of collisions between the mixture species.
A method to obtain exact solution for these equations
% transport
%coefficients of the multicomponent Fermi-liquid
was developed in \cite{Anderson1987}. Below we adapt this
formalism to the form convenient to perform the calculations for
nuclear matter.

The necessary microscopic input %for these equations
is the values of $m^*$ and the quasiparticle scattering amplitude
$\hat{T}$. We obtain $m^*$ and $\hat{T}$ within the BHF
approximation, see \cite{Shternin2013PhRvC} for details. Consider
the scattering of two nucleon species $c$ and $i$, $c+i\to c'+i'$.
We define ${\cal Q}_{ci}(P,q)=1/4(1+\delta_{ci})^{-1} \sum_{\rm spins} |\langle
ci|\hat{T}|c'i'\rangle|^2$, where all particles are placed on the
Fermi surface and the momentum and energy conservation
delta-functions are taken out. This quantity depends on the total
nucleon pair momentum $\bm{P}=\bm{p}_c+\bm{p}_i$ and the
transferred momentum $\bm{q}=\bm{p}_c'-\bm{p}_c$. In what follows
we will need the angular averages ${\cal Q}_{ci}^{(kl)}=\langle
{\cal Q}_{ci} P^k q^l\rangle $. The explicit definition for these
averages which have the meaning of  generalized transport
cross-sections can be found in the appendix in
\cite{Shternin2013PhRvC}. When the nucleon interaction is expanded
in partial waves, such representation is more convenient than the
traditional use of the Abrikosov-Khalatnikov angles
\cite{Shternin2013PhRvC}.
%\begin{equation}\label{eq:Qnm}
%  {\cal Q}^{(k,l)}_{ci}=
%  \int\limits_{|p_{Fc}-p_{Fi}|}^{p_{Fc}+p_{Fi}} {\rm d } P
%  \int\limits_{0}^{q_m(P)} \frac{P^k q^l\;{\rm d}
%  q}{\sqrt{q_m^2(P)-q^2}}{\cal Q}_{ci}(P,q),
%\end{equation}
%where $q_m(P) P^2= 4p_{Fc}^2p_{Fi}^2-(p_{Fc}^2+p_{Fi}^2-P^2)^2$.

%The averages (\ref{eq:Qnm}) can be
%seen as the generalized transport cross-sections.
%When the nucleon
%interaction is expanded in partial waves, the internal integral
%(over $q$) can be calculated analytically
%\cite{Shternin2013PhRvC}.

In transport theory, one expresses a non-equilibrium distribution
function $F(\bm{p},\epsilon)$ by introducing a linear (in external
perturbation) correction to the local equilibrium distribution
function as
\begin{equation}\label{eq:F_anzatz}
  F(\bm{p},\epsilon)=f(\epsilon)+ \tau \Psi(x)  D(\bm{p})
\frac{\partial f}{\partial
  \epsilon},
\end{equation}
where $f(\epsilon)$ is the Fermi-Dirac distribution,
$x=(\epsilon-\mu)/(k_{\rm B}T)$, $\mu$ is the chemical potential
and $D(\bm{p})$ is the anisotropic part of the driving term. For
example,  $D(\bm{p}) = \bm{v}\nabla T$ for thermal conductivity.
The transport coefficients are given then by (\ref{eq:kappa_eta})
with
\begin{equation}
  C_{\kappa/\eta} = \xi \int\limits_{-\infty}^{+\infty} {\rm d} x\, \Xi(x) \Psi(x)
  f(x) (1-f(x)).
\end{equation}
For the shear viscosity, $\Xi(x)=1$ and $\xi=1$, while for thermal
conductivity,  $\Xi(x)=x$ and $\xi=3/\pi^2$.
%\begin{equation}
% - \Xi(x) D \frac{\partial f}{\partial \epsilon} = \left(\frac{{\rm d} f}{{\rm d}
% t}\right)_{\rm coll}
%\end{equation}
The unknown functions $\Psi_c(x)$, where $c$ now enumerates
particle species, are determined from  integral equations, derived
by the linearization of the system of transport equations
\cite{Anderson1987} using (\ref{eq:F_anzatz}). Defining the
characteristic relaxation time as
\begin{equation}\label{eq:tau_c}
\tau_c = \frac{4\pi^2 \hbar^7 p_{{\rm F}c}}{m_c^* (k_{\rm B} T)^2}
{\cal N}_c,\quad {\cal N}_c=\left(\sum_{i} m_i^{*^2} {\cal
Q}_{ci}^{(00)} \right)^{-1},
\end{equation}
%where the normalization factor ${\cal N}_c=\left(\sum_{i}
%m_i^{*^2} {\cal Q}_{ci}^{(00)} \right)^{-1}$,
one obtains the system of integral equations in a simple form
\begin{equation}\label{eq:psi_eq}
  \Xi (x) f(-x)=\left(1+\frac{x^2}{\pi^2}\right) \Psi_c(x)f(-x) -
  \frac{2}{\pi^2} \int\limits_{-\infty}^{+\infty} {\rm d}\,x'\,
  f(-x') \frac{x-x'}{1-{\rm e}^{x'-x}} \sum_{i} \lambda_{ci}
  \Psi_i (x'),
\end{equation}
where all information about the quasiparticle scattering is
encapsulated in the matrix $\lambda_{ci}$. For $\kappa$ and $\eta$
problems, the matrices $\lambda_{ci}$ are expressed through
angular averages ${\cal Q}^{(kl)}_{ci}$ as
\begin{eqnarray}
\lambda^{\kappa}_{cc} &=& {\cal N}_c\sum_i m_i^{*2} \left( {\cal
Q}_{ci}^{(00)}-\frac{1}{2 p_{{\rm F}c}^2}{\cal Q}_{ci}^{(02)}\right)+\left.\lambda^\kappa_{ci}\right|_{i=c},\\
\lambda^{\kappa}_{ci}&=& \frac{\tau_i}{ p_{{\rm F}c}^2 \tau_c}
{\cal N}_c\, m_c^{*} m^{*}_i \left({\cal
Q}_{ci}^{(20)}-\left(p_{{\rm F}c}^2+p_{{\rm F}i}^2\right) {\cal
Q}_{ci}^{(00)}+\frac{1}{2}{\cal Q}_{ci}^{(02)} \right),\\
\lambda^{\eta}_{cc} &=& {\cal N}_c\sum_i m_i^{*2} \left( {\cal
Q}_{ci}^{(00)}-\frac{3}{ 2p_{{\rm F}c}^2}{\cal Q}_{ci}^{(02)}+\frac{3}{ 8p_{{\rm F}c}^4}{\cal Q}_{ci}^{(04)}\right)+\left.\lambda^\eta_{ci}\right|_{i=c},\\
\lambda^{\eta}_{ci}&=& \frac{3\tau_i}{4 p_{{\rm F}c}^4 \tau_c}
{\cal N}_c\, m_c^{*} m^{*}_i \left({\cal
Q}_{ci}^{(22)}-\left(p_{{\rm F}c}^2+p_{{\rm F}i}^2\right) {\cal
Q}_{ci}^{(02)} +\frac{1}{2}{\cal Q}_{ci}^{(04)} \right).
\end{eqnarray}
Provided the matrix $\lambda_{ci}$ is known, the solution of the
system (\ref{eq:psi_eq}) can be found numerically by iteration
method, as we do here. In \cite{Anderson1987} the general method
is developed to express the solution in form of the rapidly
converging series. The frequently used simplest variational
solution corresponds to a first term in these series (namely,
$\Psi(x)\propto \Xi(x)$). For carriers of one type, these
variational solutions give $C_\kappa^{\rm Var}=1.25
(3-\lambda^\kappa_{cc})^{-1}$ and $C_\eta^{\rm Var}=0.75
(1-\lambda^\eta_{cc})^{-1}$ \cite{Anderson1987}.
%To obtain these solutions,  the trial
%functions $\Psi(x)\propto \Xi(x)$ are used in (\ref{eq:psi_eq}).

\begin{figure}[t]
\begin{center}
   \begin{minipage}[h]{0.4\textwidth}
     \caption{ Functions $\Psi_\eta(x)$ and $\Psi_\kappa(x)/x$ for neutrons (\full) and protons (\dashed) from the exact solutions of the system (\ref{eq:psi_eq})
     for the Av18+TBF microscopic model and $n_{\rm B}=0.3$~fm$^{-3}$. Thick red
     and thin blue lines correspond to thermal conductivity and shear
     viscosity problems, respectively.
%     For the thermal conductivity the function $\Psi_\kappa (x)/x$ is plotted.
 }\label{fig:psi}
 \end{minipage}
 \hspace{0.05\textwidth}%
 \begin{minipage}{0.35\textwidth}
    \includegraphics[width=\textwidth]{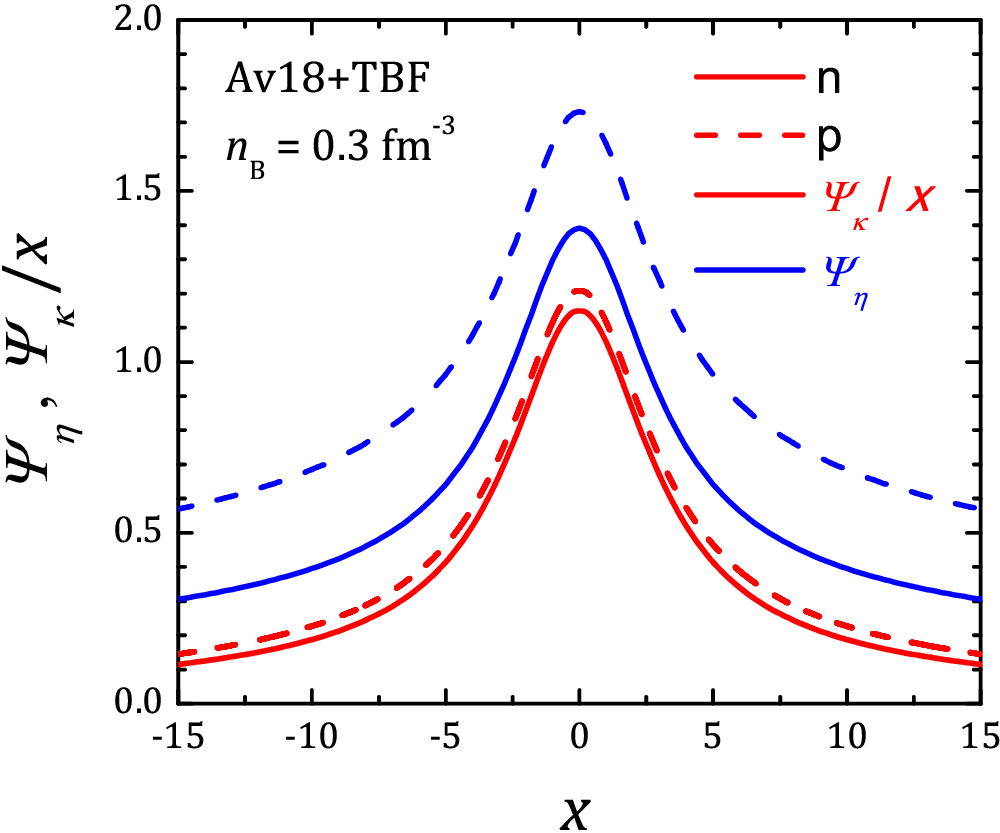}
    \end{minipage}
\end{center}
\end{figure}

\section{Results and conclusions}\label{sec:results}
As an example, consider the Av18+TBF model at the total baryon
density $n_{\rm B}=0.3$~fm$^{-3}$ and $T=10^8$~K. The proton
fraction of $\beta$-stable matter is $x_{\rm p}=0.1$, the
effective masses (in units of nucleon mass) are $m_{\rm n}^*=0.81$
and $m_{\rm p}^*=0.67$ \cite{Baldo2014PhRvC}. The neutron and
proton effective relaxation times (\ref{eq:tau_c}) are similar,
$\tau_{\rm n}=1.2\times 10^{-16}$~s and $\tau_{\rm p}=0.79\times
10^{-16}$~s, but since $n_{\rm p}$ is much smaller than $n_{\rm
n}$, the proton contribution to transport coefficients is small
[see (\ref{eq:kappa_eta})].  The exact solutions of the system
(\ref{eq:psi_eq}) are shown in figure~\ref{fig:psi} giving $C^{\rm
n}_\kappa = 0.67$ and $C^{\rm n}_\eta =1.21 $ for the neutrons.
The total value of the thermal conductivity is $\kappa=\kappa_{\rm
n}+\kappa_{\rm p} = 1.2\times
10^{23}$~erg~cm$^{-1}$~s$^{-1}$~K$^{-1}$ and the total shear
viscosity is $\eta = 2.8\times 10^{18}$~g~cm$^{-1}$~s$^{-1}$. It
is instructive to note that the influence of  $\Psi_{\rm p}(x)$ on
the equation for $\Psi_{\rm n}(x)$ via the non-diagonal terms in
(\ref{eq:psi_eq}) is negligible, since $\lambda_{\rm
nn}^\kappa=0.07\lambda_{\rm np}^\kappa$ (???).
 However, the inverse is not true,
 $\lambda_{\rm pn}^\kappa=1.39\lambda_{\rm pp}^\kappa$ and the non-equilibrium
neutron distribution strongly affects the proton one. Thus it
seems a good starting approximation to consider only the neutron
contribution to transport (but taking into account np collisions
as well, see, e.g., \cite{Shternin2013PhRvC}). In the present
example, this gives $C_\kappa^{{\rm n},\ {\rm Var}}=0.55$ and
$C_\eta^{{\rm n},\ {\rm Var}}=1.13$ where we additionally used the
simplest variational solution. Comparing this approximation with
the exact values, we get $\kappa/\kappa_{\rm n}^{{\rm Var}} =
1.36$
%. Similarly, for the shear
%viscosity,
and $\eta/\eta_{\rm n}^{{\rm Var}}=1.1$.

\begin{figure}[t]
\begin{center}
\begin{minipage}{0.37\textwidth}
\includegraphics[width=\textwidth]{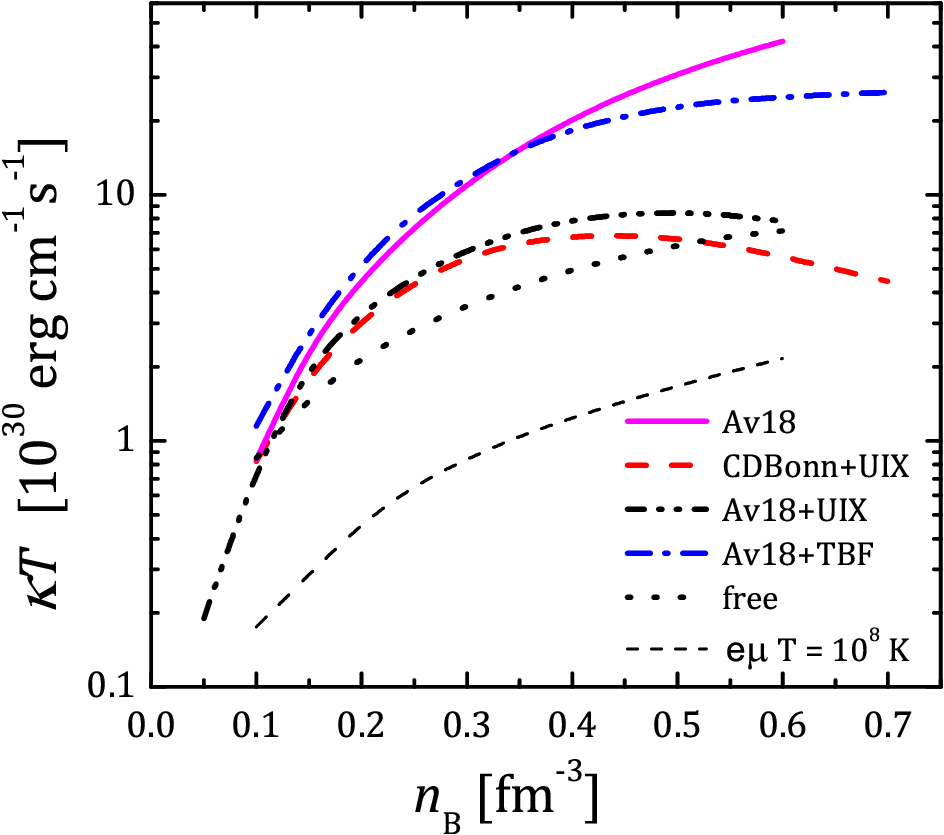}
\end{minipage}\hspace{0.1\textwidth}%
\begin{minipage}{0.39\textwidth}
\includegraphics[width=\textwidth]{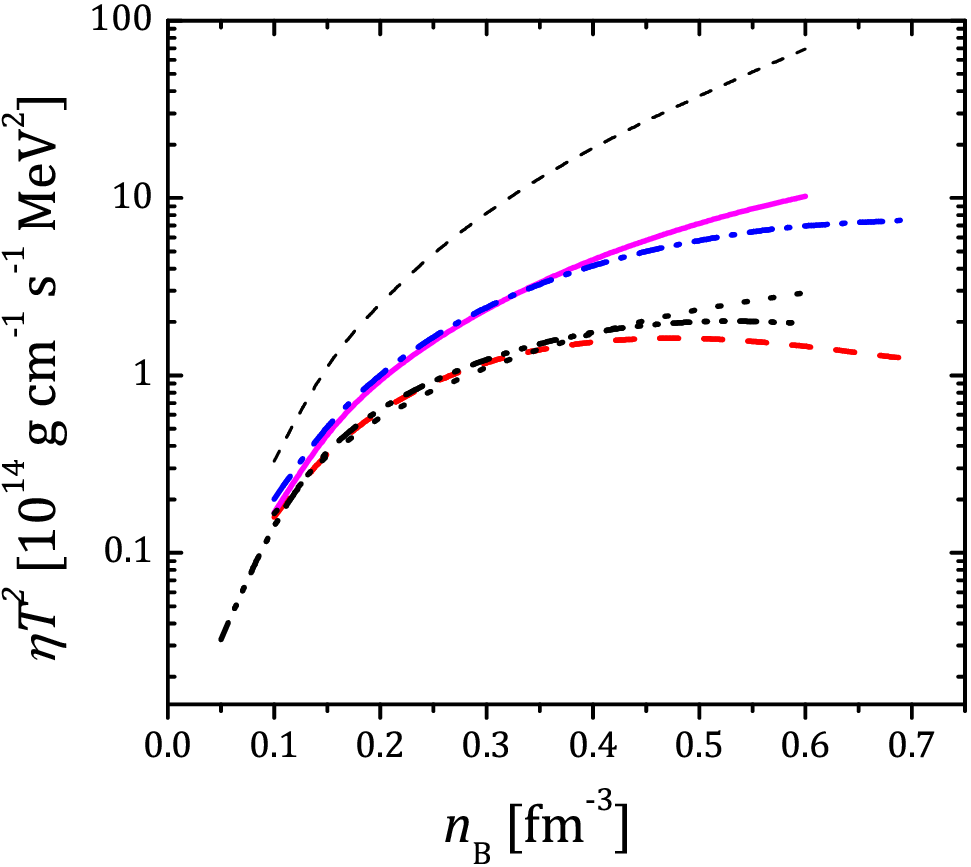}
\end{minipage}
\end{center}
\caption{\label{fig:kin} Temperature-independent combinations
$\kappa T$ ({\it left}) and $\eta T^2$ ({\it right}) as functions
of baryon number density for the models considered in the text.
Thin dashed lines show lepton contribution for $T=10^8$~K and
Av18+UIX EOS calculated according
\cite{ShterninYakovlev2007,ShterninYakovlev2008}.}
\end{figure}

In figure~\ref{fig:kin} we present the final results of our
calculations. We show the results for the Av18 potential on the
two-body level and with addition of the UIX and TBF three-body
forces. The CD-Bonn results on the two-body level give similar
values as Av18 and are not shown, while the CD-Bonn+UIX results
are shown with red dashed lines. For comparison, we show also the
results obtained using free-space scattering probability and
$m^*=1$ (dotted lines) as well as the lepton contribution (thin
dashed lines). Notice, that the lepton contribution has
non-Fermi-liquid  temperature dependence ($\eta_{\rm e\mu}\propto
T^{-5/3}$ and $\kappa_{\rm e\mu}={\rm const}(T)$)
\cite{ShterninYakovlev2007,ShterninYakovlev2008}. One clearly sees
the importance of in-medium effects. At a two-body level, both
$\eta$ and $\kappa$ in-medium values are much higher than the
results based on the free-space interaction. The inclusion of UIX
three-body force acts in the opposite direction, partially (but
not fully) due to the increase in  particles' effective masses
\cite{Shternin2013PhRvC, Baldo2014PhRvC}. The effect of the
meson-exchange three-body model TBF is much less pronounced, and
the results stay approximately at the two-body level. In the
future we plan to investigate the contribution of different
partial waves to $\kappa$ and $\eta$ calculations in order to
understand the difference in UIX and TBF effects.

The variations of $\kappa$ and $\eta$ found in our calculations
are large. Fortunately, the shear viscosity is dominated by the
lepton contribution, whose dependence on the equation of state is
mainly through the particle fractions and effective masses and may
be easily accounted for. For the thermal conductivity, the
situation is reverse. However, the precise value of the thermal
conductivity is usually not important since it is so large that
the NS core is almost isothermal. In our study we did not consider
the effects of the superfluidity that can be very important in NS
cores. This is a good project for the future.

%\section{Conclusions}\label{sec:conclusions}

%\begin{itemize}
%  \item First conclusion
%  \item Second conclusion
%\end{itemize}

%Conclusions to be concluded.

\ack The work of PSS was supported by the Russian Foundation for
Basic Research, grant 16-32-00507 mol$\_$a and the ``BASIS''
Foundation.

\section*{References}
%\bibliographystyle{iopart-num}
%\bibliography{kincore.ns2017}
\providecommand{\newblock}{}

%\begin{thebibliography}{9}
%\bibitem{iopartnum} IOP Publishing is to grateful Mark A Caprio, Center for Theoretical Physics, Yale University, for permission to include the {\tt iopart-num} \BibTeX package (version 2.0, December 21, 2006) with  this documentation. Updates and new releases of {\tt iopart-num} can be found on \verb"www.ctan.org" (CTAN).
%\end{thebibliography}

%\appendix
%\section*{Appendix}
%\setcounter{section}{1}
%\begin{equation}\label{eq:Qnm}
%  {\cal Q}^{(k,l)}_{ci}=
%  \int\limits_{|p_{Fc}-p_{Fi}|}^{p_{Fc}+p_{Fi}} {\rm d } P
%  \int\limits_{0}^{q_m(P)} \frac{P^k q^l\;{\rm d}
%  q}{\sqrt{q_m^2(P)-q^2}}{\cal Q}_{ci}(P,q),
%\end{equation}
%where $q_m(P) P^2= 4p_{Fc}^2p_{Fi}^2-(p_{Fc}^2+p_{Fi}^2-P^2)^2$.

\end{document}